\documentclass[12pt]{article}
\usepackage{amsmath,amsfonts,amssymb,amscd,latexsym,lineno,graphicx,epsfig}
\usepackage[usenames]{color}
\usepackage{lingmacros}

\newcommand{\be}{\begin{equation}}
\newcommand{\en}{\end{equation}}

\newcommand{\re}{\textrm{e}}
\newcommand{\rd}{\textrm{d}}

\newcommand{\lshad}{[\![}
\newcommand{\rshad}{]\!]}

\numberwithin{equation}{section}

\begin{document}

\begin{center}
{\large\bf On the propagation of temperature-rate waves and traveling waves in rigid conductors of the  Graffi--Franchi--Straughan type}
\\ [20pt]
{\small
 {\bf Sandra Carillo} \\
Dipartimento di Scienze di Base e Applicate \\
per l'Ingegneria, 
 Universit\`{a}   di Roma  {\textsc{La Sapienza}} ,\\
 Via Antonio Scarpa 16,  00161 Rome,  
Italy\\
\& \\
I.N.F.N. - Sezione Roma1,
Gr. IV - M.M.N.L.P.,  Rome, Italy\\
[10pt]
 {\bf Pedro M.  Jordan\footnote{Also at:  Acoustics Divivision, U.S.\ Naval Research Laboratory, Stennis Space Center, Mississippi 39529, USA.}} \\ Entropy Reversal Consultants, L.L.C.,\\ P.\ O.\ Box 0691, Abita Springs, LA 70420,  USA}
\end{center}

\begin{abstract}
We examine second-sound phenomena  in a class of rigid, thermally conducting, solids that are described by a special case of the Maxwell--Catteneo flux law. Employing both analytical and numerical methods, we examine both temperature-rate waves and thermal traveling waves in this class of thermal conductor, which have recently been termed  Graffi--Franchi--Straughan type conductors.  In the present study, the temperature-dependent nature of the thermal relaxation time, which is the distinguishing feature of this class of conductors, gives rise to a variety of nonlinear effects; in particular, finite-time temperature-rate wave blow-up and  temperature traveling waveforms which exhibit a ``tongue''.  The presentation concludes with a discussion of  possible follow-on studies.

\end{abstract}

{\bf keyword}
\noindent
 Graffi--Franchi--Straughan conductors;  Maxwell--Cattaneo law;   temperature-rate waves;  traveling wave solutions;  Lambert $W$-function


\section{Introduction}\label{sect:Intro}

Absent the presence of source terms, the most general form of the system that 
describes  the propagation of second-sound (i.e., thermal waves)   in the `class' of 
rigid, homogeneous and isotropic, solids that the present authors have termed 
\emph{Graffi--Franchi--Straughan} (GFS) conductors~\cite{SP16} reads
\begin{subequations}\label{Sys:GFS_3D}
\begin{equation}\label{eq:_MC_GFS_3D}
{\bf q} + \alpha \vartheta\mathcal{K}(\vartheta)   {\bf q}_t = -\mathcal{K}(\vartheta)\boldsymbol{\nabla}\vartheta,
\end{equation}   
\begin{equation}\label{eq_Energy_3D}
\rho c_{\rm v}(\vartheta)\vartheta_t +\boldsymbol{\nabla\cdot}\, {\bf q} = 0,
\end{equation}
\end{subequations}
where $\vartheta (>0)$ denotes the  absolute temperature, ${\bf q}$ is the heat flux vector,  $c_{\rm v}(\vartheta)(>0)$ is the constant-volume specific heat~\cite{Blackman55}, $\mathcal{K}(\vartheta)(>0)$ is the thermal 
conductivity,   $\alpha$ is a positive constant that carries (SI) units of ${\rm 
m}\cdot {\rm sec}/{\rm W}$,   and  $\rho(>0)$ is the (constant)   mass density of the conductor.  Eq.~\eqref{eq:_MC_GFS_3D}, we observe, is  a special case of the  Maxwell--Cattaneo (MC) law, i.e., a special case of the flux constitutive relation~\cite{CN88,JP89,JCL10,MR93,OS09,S11}
\begin{equation}\label{eq:_MC_3D}
{\bf q} + \tau(\vartheta)  {\bf q}_t = -\mathcal{K}(\vartheta)\boldsymbol{\nabla}\vartheta;
\end{equation}   
specifically, the former follows from the latter on setting $\tau(\vartheta)=\tau_{\rm GFS}(\vartheta)$, where
\be\label{eq:tau_GFS}
\tau_{\rm GFS}(\vartheta) := \alpha \mathcal{K}(\vartheta)  \vartheta, 
\en
which is the distinguishing feature of GFS  conductors.   Here,  $\tau(\vartheta)(>0)$ denotes the 
relaxation time for phonon processes that are dissipative because they do not 
conserve phonon momentum~\cite{CN88}.   

Recalling  arguments from an earlier (unpublished) contribution by Graffi, Franchi and Straughan~\cite{fs94}, in 1994,  used the derivation of Eq.~\eqref{eq:_MC_GFS_3D} to illustrate a theoretically-motivated  means by which the temperature dependence of $\tau$ may arise.  In particular, they noted that Eq.~\eqref{eq:_MC_GFS_3D} identically satisfies the following generalization of the Clausius--Duhem inequality:
\be
 (\alpha \vartheta  {\bf q}_t +\boldsymbol{\nabla}\vartheta) \boldsymbol{\cdot} {\bf q} \leq 0,
\en
which Franchi and Straughan~\cite[p.~728]{fs94} attribute to Graffi; see also Franchi~\cite{Franchi85} and Straughan~\cite[\S 1.2]{S11}.  As alluded to above,  however, GFS conductors must at present be regarded as hypothetical constructs since, to the best of our knowledge,  the literature does \emph{not} contain any examples of  actual solids wherein {\bf q} is described by Eq.~\eqref{eq:_MC_GFS_3D}. 

Nevertheless, GFS conductors exhibit a number of interesting mathematical properties that, from our perspective, make them worthy of investigation.   For example, the GFS form of $\tau$ plays a critical role in establishing   the following:

\begin{itemize}
\item  The empirically based relation for second-sound in  rigid solids
\be\label{eq:tau_CO}
\tau(\vartheta) = \frac{\mathcal{K}(\vartheta)(A_{0}+B_{0}\vartheta^{n})}{\rho c_{\rm v}(\vartheta)},
\en
where $n$, $A_{0}$, and $B_{0}$ are fitting parameters, has been shown to be applicable to NaF, for $10.0 \leq \vartheta \leq 18.5\,$K, and Bi, for  $1.4 \leq \vartheta \leq 4.0\,$K; see, e.g., Refs.~\cite{ColLai94,CN88} and those cited therein.  On comparing with Sys.~\eqref{Sys:GFS_3D}, it is easily seen that setting  $n=1$, $A_{0}=0$,  and $B_{0}=\alpha \rho c_{\rm v}(\vartheta)$ reduces  Eq.~\eqref{eq:tau_CO} to Eq.~\eqref{eq:tau_GFS}.   However, the fact  that  $B_{0}$ is a constant necessitates the additional requirement $c_{\rm v}(\vartheta) :=$~const. This is  true, as has long been known, in the case of many real solids when $\vartheta \gg \vartheta_{\rm D}$, where    $\vartheta_{\rm D}$ denotes  the  Debye temperature of the solid in question; see, e.g., Ref.~\cite[\S 2]{Blackman55}.  From a strictly theoretical standpoint, then,  Eq.~\eqref{eq:tau_CO} also applies  to GFS conductors in the  \emph{high-temperature} regime, i.e., under conditions yielding $c_{\rm v}(\vartheta) :=$~const.

\item When $\tau(\vartheta)$ is given by Eq.~\eqref{eq:tau_GFS} and $c_{\rm v}(\vartheta)$ and $\mathcal{K}(\vartheta)$ are both taken to be constant, the flux relation under Morro--Ruggeri (MR) theory~\cite{MR88} reduces to its  simplest possible special case that still exhibits explicit dependence on $\vartheta_{t}$; i.e., the simplest possible special case in which Ref.~\cite[Eq.~(48)]{MR88}  does \emph{not} degenerate into a particular case of 
Eq.~\eqref{eq:_MC_3D}~(above).

\item  When $\tau(\vartheta)$ is given by Eq.~\eqref{eq:tau_GFS}, $\tilde{e}=\tilde{e}(\vartheta, {\bf q})$, the generalized expression for the internal energy density under Coleman--Fabrizio--Own (CFO) theory~\cite{CFO82} (see also Refs.~\cite{ColLai94,CN88,MR88}), reduces to it simplest possible special case that still exhibits explicit dependence on ${\bf q}$; i.e., the simplest possible special case in which $\tilde{e}(\vartheta, {\bf q})$  does \emph{not} degenerate into the classical expression for the internal energy density of a rigid solid.


\end{itemize}

The primary aim of this communication is to present numerical simulations of the  
second-sound phenomena  that the present authors examined in 
Ref.~\cite{SP16} using only analytical methods.  In particular, we 
 simulate both temperature-rate waves and traveling waves predicted by the 
 following  special case of  Sys.~\eqref{Sys:GFS_3D}:
\begin{subequations}\label{Sys:GFS}
\begin{equation}\label{eq:_MC_GFS}
q + \alpha K \vartheta q_t = -K\vartheta_x,
\end{equation}   
\begin{equation}\label{eq_Energy}
\rho c_{\rm p}\vartheta_t + q_x = 0.
\end{equation}
\end{subequations}
As in Ref.~\cite{SP16}, we have assumed $\mathcal{K}(\vartheta):= K$, where the constant $K$ is the value of the thermal conductivity at some reference temperature;  we have taken  $c_{\rm v}(\vartheta)$ to be a constant ($\Rightarrow \vartheta \gg \vartheta_{\rm D}$), but have also made use of the fact that $c_{\rm v} = c_{\rm p}$ under the rigid solid idealization\footnote{For details on the justification/rational behind the use of the approximation $c_{\rm v} \approx  c_{\rm p}$ when modeling the flow of heat in real solids,  see, e.g., Refs.~\cite{CJ59, FW80, Powers19}.}, where $c_{\rm p}$ denotes the constant-pressure specific heat; and  we have confined our attention  to  one-dimensional (1D) heat flow along the $x$-axis, a propagation geometry which renders  $\vartheta =\vartheta (x,t)$ and ${\bf q} = (q(x,t),0,0)$.

To this end, the present article is organized as follows. In 
Sect.~\ref{sec:accwavs}, a review of the temperature-rate wave analysis  carried 
out in Ref.~\cite{SP16} is presented.    In Sect.~\ref{sec:numerical},  numerical 
simulations of temperature-rate waves  are performed and results obtained are 
compared with our analytical findings. Then, in Sect.~\ref{sec:tws}, a traveling 
wave analysis  of Sys.~\eqref{Sys:GFS} is performed and a two of the 
resulting solution profiles are studied numerically. And lastly, in 
Sect.~\ref{sec:concl}, connections to other works are discussed and possible 
follow-on studies are noted.  

\section{Temperature-rate waves: Analytical results}\label{sec:accwavs}
\subsection{Brief history and related works}
By a temperature-rate wave\footnote{Also known as a temperature-rate 
discontinuity wave and  a discontinuity (or acceleration) wave; see, e.g., 
Refs.~\cite{Morro06} and~\cite[\S 7.4]{MR93}, respectively.} we mean a singular 
surface, i.e., a wavefront, across which the first derivatives of the temperature 
field suffer a jump discontinuity; see, e.g., Refs.~\cite{Morro06,S11}.  What 
makes these waves so interesting 
is the fact that, under certain conditions, the jump amplitude can  exhibit {\it 
finite-time blow-up}, even when the imposed thermal disturbance is continuous. 
Today, it is generally accepted that temperature-rate wave amplitude  blow-up 
signals the formation  of a {\it thermal shock}~\cite{S11}, i.e., a propagating surface across which  $\vartheta$ itself suffers a jump.  

\subsection{Mathematical preliminaries: Characteristic speed}\label{sect:MathPrelim}
Letting $\kappa =  K/(\rho c_{\rm p})$ denote the \emph{thermal diffusivity}~\cite{FW80}, 
we begin this subsection by recasting Sys.~\eqref{Sys:GFS} in (equivalent) matrix form, 
specifically, as
\begin{equation}
\begin{pmatrix}
\vartheta \\
q
\end{pmatrix}_t
+
\mathsf{A}
\begin{pmatrix}
\vartheta \\
q 
\end{pmatrix}_x=
- (\alpha K \vartheta)^{-1}\begin{pmatrix}
0 \\
q
\end{pmatrix}\!, \quad
\textrm{where} \quad
\mathsf{A}=\begin{pmatrix}
0 & \kappa/K   \\
1/(\alpha  \vartheta) & 0 
\end{pmatrix}\!.
\label{Sys:GFS_Matrix}
\end{equation}
The eigenvalues, $\mu_{1,2}$, of the  coefficient matrix satisfy  the characteristic equation $
\det(\mathsf{A}-\mu \mathsf{I}_2)=0$, where  $\mathsf{I}_2$ denotes the $2\times 
2$ identity matrix; thus, $\mu_{1,2}=\pm U
(\vartheta)$, where    the characteristic speed of second-sound under Sys.~\eqref{Sys:GFS_Matrix} is
\be\label{eq:U_dim}
U(\vartheta)=\sqrt{\frac{\kappa}{\alpha K \vartheta}},
\en
and  the  characteristics of Sys.~\eqref{Sys:GFS_Matrix} are defined by $
{\rm d} x/{\rm d}t=\pm U(\vartheta)$.  Since $\mu_{1,2}\in \mathbb{R}$  and unequal, it follows that this (quasilinear) 
system is a \emph{strictly hyperbolic}~\cite{Logan08} one.  Therefore,  solutions of Sys.~\eqref{Sys:GFS_Matrix} satisfy the requirements of causality \emph{provided}  $U(\vartheta)$ is bounded, where  from Eq.~\eqref{eq:U_dim} we see that  $U(\vartheta) \to \infty$ as $\vartheta \to 0$.

Since $\vartheta \gg \vartheta_{\rm D}$ is one of the assumptions on which Sys.~\eqref{Sys:GFS} is based,   the  breakdown of our model as $\vartheta \to 0$ does not pose a difficulty for us.  To gain deeper insight into the behavior of $U
(\vartheta)$, it is instructive to consider finite- and small-amplitude thermal disturbances.  Under the weakly-nonlinear 
and linear approximations Eq.~\eqref{eq:U_dim} becomes
\be\label{eq:U_wnl}
U_{\rm wnl}(\vartheta) = \sqrt{\frac{\kappa}{\alpha K \vartheta_{0}}}
\left[1-\frac{1}{2}\left( \frac{\vartheta-\vartheta_{0}}{\vartheta_{0}}\right) \right] 
\qquad (|\vartheta-\vartheta_{0}| \ll \vartheta_{0}),
\en
where in this study $\vartheta_0$ denotes the initial temperature of the 
conductor, and
\be
U_{0}=\sqrt{\frac{\kappa}{\alpha K \vartheta_{0}}},
\en
respectively.  Here, we see that $U_{\rm wnl}(\vartheta) > U
_{0}$, when $\vartheta_{0} > \vartheta$, while  $U_{\rm wnl}
(\vartheta) < U_{0}$, when $\vartheta_{0} < \vartheta$.  This, we 
observe, is the \emph{opposite} of the behavior exhibited under the (bi-directional) 
model equations of classical acoustics; see, e.g., Ref.~\cite[\S 4($a$)]
{j05}, and note that  $\wp$, the thermodynamic pressure in Ref.~\cite{j05}, 
corresponds to $\vartheta$.  

Lastly, to simplify the forthcoming temperature-rate wave analysis, we now 
introduce the following non-dimensional variables:
\begin{equation}
\theta=\vartheta/\vartheta_{0},\quad x^\circ=x/L,\quad t^\circ=t(\kappa /L^2), 
\quad q^\circ= q(L/(K\vartheta_0)),
\label{ND_var}
\end{equation}
 where $L$ denotes the conductor's  thickness, and recast 
Sys.~\eqref{Sys:GFS_Matrix} in non-dimensional form, viz.:
\begin{equation}
\begin{pmatrix}
\theta \\
q
\end{pmatrix}_t
+
\begin{pmatrix}
0 & 1 \\
1/(\lambda \theta) & 0 
\end{pmatrix}
\begin{pmatrix}
\theta \\
q 
\end{pmatrix}_x=
-(\lambda \theta)^{-1} \begin{pmatrix}
0 \\
q
\end{pmatrix}\!.
\label{System}
\end{equation}
 where, for convenience, we have set
\be\label{eq:lambda}
\lambda := \alpha \kappa K\vartheta_{0}/ L^{2},
\en
all superscript circles have been omitted but should remain understood, and we note for later reference that  
$\mathcal{C}(\theta)=(\lambda \theta)^{-1/2} $ is the \emph{non-dimensional} form of $U(\vartheta)$.

\subsection{Formulation}\label{sect:slab}
Now consider a   rigid conducting slab,  whose (normalized) thickness is unity, 
wherein the temperature and  heat flux are described by Sys.~\eqref{System}.  
We suppose the slab is stationary and that, initially, $q=0$ and the slab is at a 
uniform temperature $\theta(x,0)= 1$ (i.e., $\vartheta(x,0) =\vartheta_{0}$) 
throughout.   Beginning at time $t=0+$, let a temperature pulse  of the form
\begin{equation}
\theta(0,t)= 1+ H_{\rm p}(t,t_{\rm w})\psi(t), \quad \textrm{where} \quad H_{\rm p}
(t,t_{\rm w}) := H(t)-H(t-t_{\rm w}),
\label{BC}
\end{equation}
be applied to the boundary $x=0$, while the boundary $x=1$ is held at 
temperature $\theta(1,t)=1$.  Here, the pulse duration (or width) $t_{\rm w}(>0)$ 
is a constant; the amplitude function $|\psi(t)| \in (0,1)$ is assumed to be  
continuously differentiable, nonzero on the interval $t \in (0,t_{\rm w})$, and such 
that $\psi(0)=0$ but $\psi_{t}(0)\neq 0$; and $H(\cdot)$ denotes the Heaviside 
unit step function. 

Since Sys.~\eqref{System} is  strictly hyperbolic, a planar 
temperature wavefront $x=\Sigma (t)$, across which $\lshad \theta\rshad= 
\lshad q\rshad=0$, but $\lshad \theta_t\rshad \neq 0$,  begins propagating  from 
the boundary $x=0$, along the positive 
$x$-axis, with speed $\mathcal{C}(\theta^{+})=\lambda^{-1/2}$ relative to the slab, where we observe that $\theta^{+}=1$ under the present formulation.  Here, employing the 
standard notation of singular surface theory,   $\lshad \mathfrak{F}\rshad := 
\mathfrak{F}^{-}-\mathfrak{F}^{+}$ denotes the amplitude of the jump  in the 
value of the function $\mathfrak{F}=\mathfrak{F}(x,t)$ across $\Sigma(t)$, where 
$\mathfrak{F}^{\pm}\equiv \lim_{x\to \Sigma (t)^{\pm}}\mathfrak{F}(x,t)$ are 
assumed to exist, and $\pm$ superscripts correspond to the regions ahead of 
and behind $\Sigma$, respectively.  Since, under this formulation, $\theta_t$   suffers a jump discontinuity across it, the  surface $\Sigma$ is clearly a temperature-rate wave.

\subsection{Amplitude evolution}\label{sect:accel}

Observing now that $\lshad \theta_t\rshad$ is, at most, a function of only $t$, 
and referring the reader to Refs.~\cite{Bland88,S11} for  details, it is a 
straightforward matter to show that $\Sigma(t)=c_{0}t+x_0$, where we have set $c_{0} := \lambda^{-1/2}$ for convenience and $x=x_0$ is the location of $\Sigma$ at $t=0$, and that the jump in $\theta_t$  
satisfies the Bernoulli equation
\begin{equation}
2\frac{\mathfrak{d} a}{\mathfrak{d} t}=- a (\lambda^{-1} + a).
\label{Ber_Eq}
\end{equation}
Here, use has been made of 
\be\label{eq:Hada}
\frac{\mathfrak{d} \lshad \mathfrak{F} \rshad}{\mathfrak{d} t} = \lshad 
\mathfrak{F}_{t} \rshad + c_{0}\lshad \mathfrak{F}_{z} \rshad,
\en
which is usually referred to as the \emph{kinematic condition of compatibility}
\footnote{See Ref.~\cite[\S 4.1]{S11} and those cited therein; see also 
Bland~\cite[\S 6.9]{Bland88}, who refers to this relation as `Hadamard's 
lemma'.},
where $\mathfrak{d}/\mathfrak{d} t$, the 1D displacement derivative,  gives the 
time-rate-of-change measured by an observer traveling with $\Sigma$; we have 
set $a(t) :=\lshad \theta_t \rshad$  for convenience; and it should be noted that, 
since the slab's initial temperature was assumed to be constant, we took
$\theta_t^{+}=0$ in deriving Eq.~\eqref{Ber_Eq}. 

Making use of the  substitution 
$a=1/\mathfrak{a}$, 
Eq.~\eqref{Ber_Eq} is  transformed into a linear ODE, which is easily  integrated; its exact solution can be expressed as
\begin{equation}
a(t)=-|\alpha^* |\left\{1-\left[1+\frac{|\alpha^*|}{a(0)}\right]\exp(\tfrac{1}{2} t/
\lambda)\right\}^{-1},
\label{Jump_Amp}
\end{equation}
where the  (negative) constant $\alpha^*$, known as the \emph{critical 
amplitude}, is given by
\begin{equation}
\alpha^* :=  -\lambda^{-1}. 
\label{Alpha}
\end{equation}
According to Eq.~\eqref{Jump_Amp}, $a(t)$ can evolve in any one of the 
following four ways: 
\begin{enumerate}
\item[(i)] If $a(0)>0$,  then $a(t)\in (0,a(0))$ for $t>0$ and $a(t)\to 0$ from above 
as $t\to \infty$.
\item[(ii)] If $a(0)<0$ and $|a(0)|<|\alpha^*|$,  then $a(t)\in (a(0),0)$ for $t>0$ 
and $a(t)\to 0$ from below  as $t\to \infty$.
\item[(iii)] If $a(0)=\alpha^*$, then $a(t)=\alpha^*$ for all $t\geq 0$. 
\item[(iv)] If $a(0)<0$ and $|a(0)|>|\alpha^*|$, then $a(t) <a(0)$ for $t>0$ and, 
moreover, $|a(t)| \to \infty$ as $t \to t_{\infty}$, where 
\begin{equation}
t_{\infty}=2\lambda \ln\left[ \frac{a(0)}{a(0)+|\alpha^*|}\right]\qquad (0< t_{\infty} 
<\infty).
\label{T_infty}
\end{equation}
\end{enumerate}

 \subsection{Stability results}
While we have obtained the exact solution of Eq.~\eqref{Ber_Eq}, it is 
nevertheless instructive to investigate the steady-state behavior of $a(t)$  using 
qualitative methods; i.e., to examine the stability characteristics of the 
equilibrium solutions $\bar{a}=\{0, -|\alpha^*| \}$, which of course correspond to 
the roots of the quadratic equation $-a(|\alpha^*|+a)=0$.

    As a  phase plane analysis  reveals,   $\bar{a}=-|\alpha^*|$ is always unstable 
while  $\bar{a}=0$ is always stable.  This  means that 
a bifurcation does \emph{not} occur in the case of Eq.~\eqref{Ber_Eq}; i.e., there is 
no  \emph{interchange} of stability between the two equilibria of this ODE. 
The instability of $\bar{a}=-|\alpha^*|$ also means that  the  constant solution  in 
Case~(iii) is   unstable as well; i.e., any discrepancy, however small, in 
achieving $a(0)=\alpha^*$ will yield either Case~(ii) or~(iv).

\section{Temperature-rate waves: Numerical results}\label{sec:numerical}
While interesting and useful, temperature-rate wave results do not provide any 
information on the behavior of the temperature field {\it behind}  $\Sigma$  

Hence, to explore this aspect of the GFS model, and to illustrate the most 
important findings of Sect.~\ref{sect:accel},  we now turn to computational 
methods. In this section, we present a series of numerical simulations based on 
the slab initial-boundary value problem~(IBVP) formulated in 
Sect.~\ref{sect:slab}, which we now express as:   
\begin{subequations}
\label{IBVP_V}
\begin{equation}
V_t+c_{0}^{-2} (1+V_x)V_{tt}-V_{xx}=0,\qquad (x,t)\in(0,
1)\times (-\infty, t_{\rm r});\label{ND_NLheateq}
\end{equation}
\begin{equation} 
V_x(0, t)= \delta H_{\rm p}(t,t_{\rm r})\sin(\pi t), \quad V_x(1,t)=0, 
\quad t\in (-\infty, t_{\rm r});
\end{equation}
\begin{equation}
V(x,0)=0,\qquad V_t(x,0)=0,\qquad x\in(0,1).
\end{equation}
\end{subequations}
Here, so that Sys.~\eqref{System} could be recast as a single PDE, we have 
introduced
\begin{equation}
V: \Omega\subset \mathbb{R}^2 \mapsto \mathbb{R} \qquad s.t.\ \qquad V_x = -1+\theta,  
\end{equation}
where $\Omega =\{(x,t): 0 < x < 1, -\infty < t < t_{\rm r}\}$.
Furthermore, $\psi(t) = \delta \sin(\pi t)$, where
 $|\delta| \in (0,1)$ is a constant; $t_{\rm w}=t_{\rm r}$, where   $t_{\rm r}=1/c_0$ 
is the time required for 
 $\Sigma$ to complete its initial transit of the interval $0<x<1$ (i.e., $
\Sigma(t_{\rm r})=1$); and of course $x_0=0$.

In the case of IBVP~\eqref{IBVP_V}  the temperature-rate wave amplitude  expression, i.e., Eq.~\eqref{Jump_Amp}, 
becomes 
\begin{equation}
\lshad V_{xx}\rshad=-c_0^{-1}\lshad V_{xt}\rshad=\frac{|\alpha^*|}{c_{0}} \left\{1-
\left[1+\frac{|\alpha^*|}{\delta \pi}\right]\exp(\tfrac{1}{2}c_{0}^{2}t)\right\}^{-1}, 
\label{Amp_IBVP}
\end{equation}
where the jump  $\lshad V_{xx} \rshad$ was determined using the expression for $
\lshad V_{xt} \rshad$,  the fact that $\lshad V_{x} \rshad =0$, and the $\lshad \mathfrak{F} \rshad =0$ special case of 
Eq.~\eqref{eq:Hada}, while the expression for
the blow-up time, i.e., Eq.~\eqref{T_infty}, assumes the  form
\begin{equation}
t_{\infty}=2\lambda \ln\left[ \frac{\delta \pi}{\delta \pi + |\alpha^*|}\right]\!.
\label{Tinfty_IBVP}
\end{equation}
Also,  it should be noted that the amplitude of the jump in the 
time derivative of the boundary condition at $x=0$ across the plane $t=0$  is $\lshad 
V_{xt}\rshad |_{t=0} = a(0)= \delta \pi$.

Since an exact analytical solution does not appear to be possible, we turn 
to the calculus of finite differences and introduce the mesh points $(x_m, t_k)$, 
where $x_m=m(\Delta x)$ for each $m=-1,0,1,\ldots, M+1$  and $t_k=k (\Delta t)
$  for each $k=0,1,2,\ldots, N$.  Here,  the spatial- and temporal-step sizes are 
defined as $\Delta x=1/M$ and $\Delta t=T/N$, respectively, where $M(\geq 2)$ 
and $N(\geq 2)$ are  integers and $T\in (0, t_{\rm r})$ is the right-hand  
endpoint\footnote{Of course, if $t_{\infty} \in (0, t_{\rm r})$, then the restriction on 
$T$ becomes $T \in (0, t_{\infty})$.} of the temporal interval over which the 
solution of IBVP~\eqref{IBVP_V} shall be computed.

With our (1D) mesh established, and guided by the treatment of similar equations 
presented in Refs.~\cite{bsj08,j05}, we construct the following simple 
discretization of Eq.~\eqref{ND_NLheateq}:
\begin{multline}
\frac{V_m^{k+1}- V_m^{k-1}}{2(\Delta t)}+c_{0}^{-2}\left(1+ \frac{V_{m+1}^{k}- 
V_{m-1}^{k}}{2(\Delta x)} \right)\!\left(\frac{V_{m}^{k+1}- 2V_{m}^{k}+V_{m}^{k-1}}
{(\Delta t)^2}\right)\\
\,-\,\frac{V_{m+1}^{k}- 2V_{m}^{k}+V_{m-1}^{k}}{(\Delta x)^2}=0,
\label{FDS_V}
\end{multline}
where $V_m^k \approx V(x_m,t_k)$. 
On setting $R = (\Delta t)/(\Delta x)$ and then solving  for  $V_{m}^{k+1}$, the 
most advanced time-step approximation, we obtain the (explicit) finite difference scheme (FDS)
\begin{multline}
V_{m}^{k+1} = \Bigg{\{}\tfrac{1}{2}(\Delta t)+c_{0}^{-2} \Bigg{[} 1+\frac{V_{m+1}
^{k}- V_{m-1}^{k}}{2(\Delta x)} \Bigg{]}\Bigg{\}}^{-1}\\
\times\Bigg{\{} R^2(V_{m+1}^{k}- 2V_{m}^{k}+V_{m-1}^{k})+\tfrac{1}{2}(\Delta 
t)V_{m}^{k-1}\\
+c_{0}^{-2}(2V_{m}^{k}-V_{m}^{k-1})\Bigg{[}1+\frac{V_{m+1}^{k}- V_{m-1}^{k}}
{2(\Delta x)}\Bigg{]}\Bigg{\}},
\label{FDS2}
\end{multline}
which holds for each $m=0,1,2,\ldots, M$ and $k=1,2,3\ldots, N-1$. In turn, 
discretization of the boundary conditions gives
	\begin{equation}\label{BCs}
	V_{-1}^{k} = V_{1}^{k}-2\delta (\Delta x)\sin(\pi t_k), \qquad V_{M+1}^{k} = 
	V_{M-1}^{k} \quad (k = 1,2,3,\ldots, N),
	\end{equation}
where we note our use of the \emph{ghost points}\footnote{A numerical device 
that allows us to discretize the Neumann boundary conditions of our problem using  centered-difference quotations, i.e.,  consistent with how the spatial derivatives in Eq.~\eqref{ND_NLheateq} are discretized in our finite difference scheme; see, e.g., Ref.~\cite{Thomas95}.} $m=-1, M+1$, while the initial conditions become
\begin{equation}\label{ICs}
	V_{m}^{0} = 0, \qquad V_{m}^{1} = V_{m}^{0} \qquad (m = -1,0,1,\ldots, 
M+1).
	\end{equation}

In Figs.~1--3 we have presented temperature profile plots corresponding to 
Cases~(i), (iii), and~(iv), respectively. These time-sequence plots depict the 
evolution of the $V_{x}$ vs.\ $x$ solution profile under IBVP~\eqref{IBVP_V}, with $V_{x}$ normalized by $\delta$, during $\Sigma$'s initial 
transit of the slab. The curves shown in solid black   were produced  from data 
sets computed by a simple 
algorithm which implemented FDS~\eqref{FDS2} 
on a desktop computer running {\sc Mathematica} (ver.~11.2). Interpolations 
between the points were then accomplished using the cubic interpolation routine 
that is a built-in part of this software package. The red broken lines, which were 
generated from Eq.~\eqref{Amp_IBVP}, have been included to illustrate the 
behavior of the temperature-rate wave amplitudes; as $V_{xx}^{+}=0$ under IBVP~\eqref{IBVP_V}, the slopes of these lines give the values 
of $V_{xx}^{-}$, at their points of tangency to the solution profiles,  at the indicated times.  

For consistency across these three figures, and ease of computation, we
have selected the common values $\lambda=1.2$  ($\Rightarrow c_{0} \approx 0.9129$, $t_{\rm r}\approx 1.0954$) and $M=2500$,  $N = 5000$, $T=1$ ($\Rightarrow R=1/2$).  It should be noted that of those we tested, with  $R=1/2$ fixed,   $\lambda=1.2$ was the smallest value of $\lambda$  for which both FDS~\eqref{FDS2} was  numerically stable  and we could place, in the case of Fig.~3,  $x_{\infty}=c_{0}t_{\infty}$  very close to, but  to the left of, the boundary $x=1$. The values of $M$ and $N$ selected were based on a heuristic search to find the smallest such values, subject to $R=1/2$, that  accurately captured the manifestation of the temperature-rate wave on the temperature profile in the last frame of Fig.~3.

In Fig.~1 we observe, as predicted in Case~(i),  the slope of the profile at the 
wavefront decreasing to zero, as $t \to \infty$, when $\delta>0$. In Fig.~2 we 
see, as predicted in Case~(iii), the slope of the profile at the wavefront remaining 
\emph{constant}; specifically, $V_{xx}^{-} = \lshad V_{xx}\rshad = -1/\lambda$. In contrast, Fig.~3\footnote{Note that in Fig.~3, $T < t_{\infty} < t_{\rm r}$.}, 
which captures  approximately $87\%$ of the `lifetime' of $\Sigma$, clearly 
illustrates  the exponential increase in $|\lshad V_{xx}\rshad |$ as $t\to t_{\infty}$, 
as predicated in Case~(iv). In particular, the last frame of Fig.~3 shows the slope 
on the leading side of our solution profile becoming nearly vertical at the wavefront, strongly 
suggesting that a thermal shock is about to form.  
\begin{figure}[b!]
\vspace*{-20mm} 
\epsfig{file=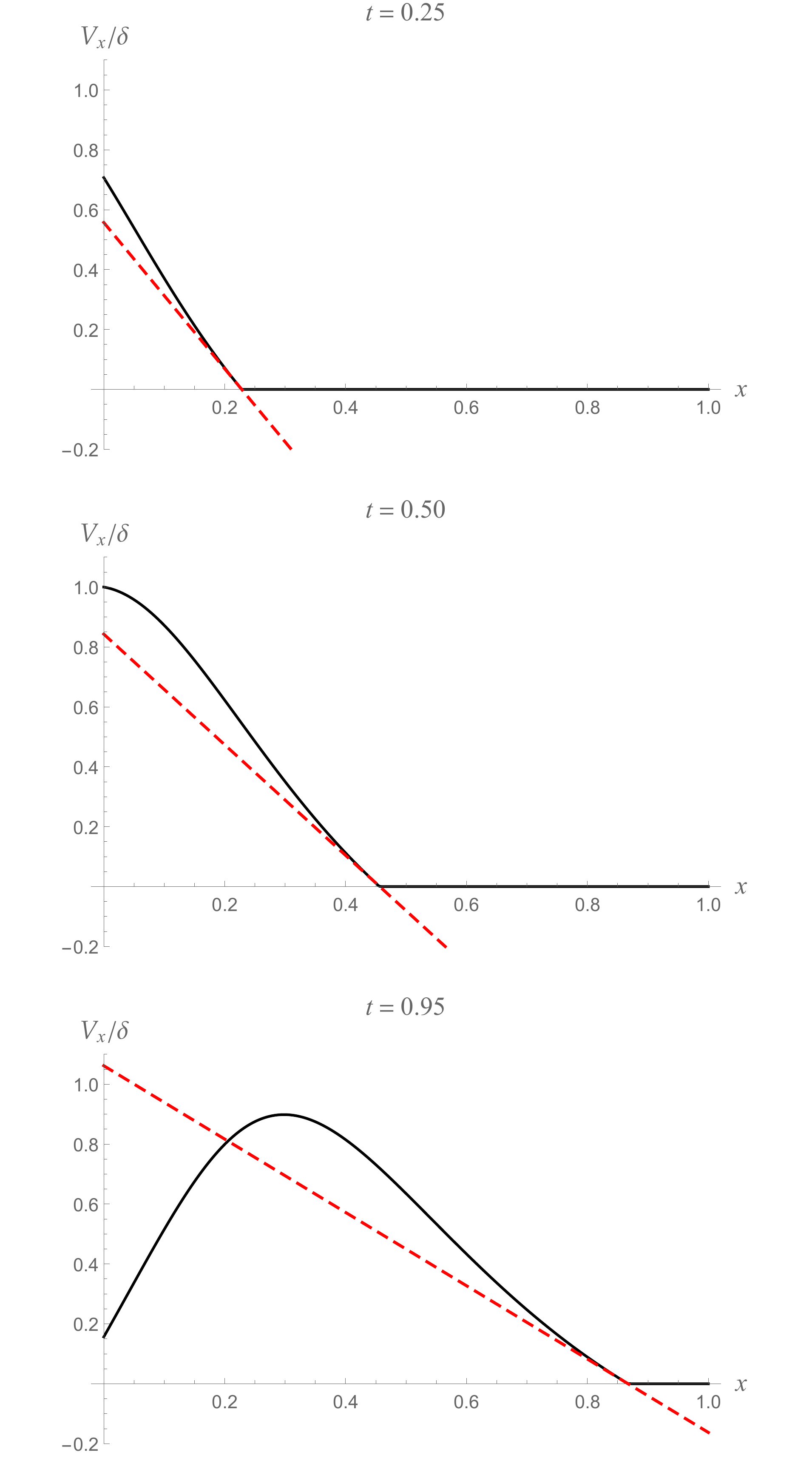, width=110mm} 
\caption{\small{$V_{x}/\delta$ vs.\ $x$ corresponding to Case~(i)  using $\lambda =1.2$ and $\delta = 0.724$, for which  $t_{\infty}, x_{\infty} < 0$.   Black solid curves: Numerically generated  
profiles using FDS~\eqref{FDS2}.  Red broken lines: Tangents at $x=\Sigma(t)$ generated using Eq.~\eqref{Amp_IBVP}.}}
\label{fig_case_i}  
\end{figure}
\begin{figure}[b!]
\vspace*{-20mm} 
\epsfig{file=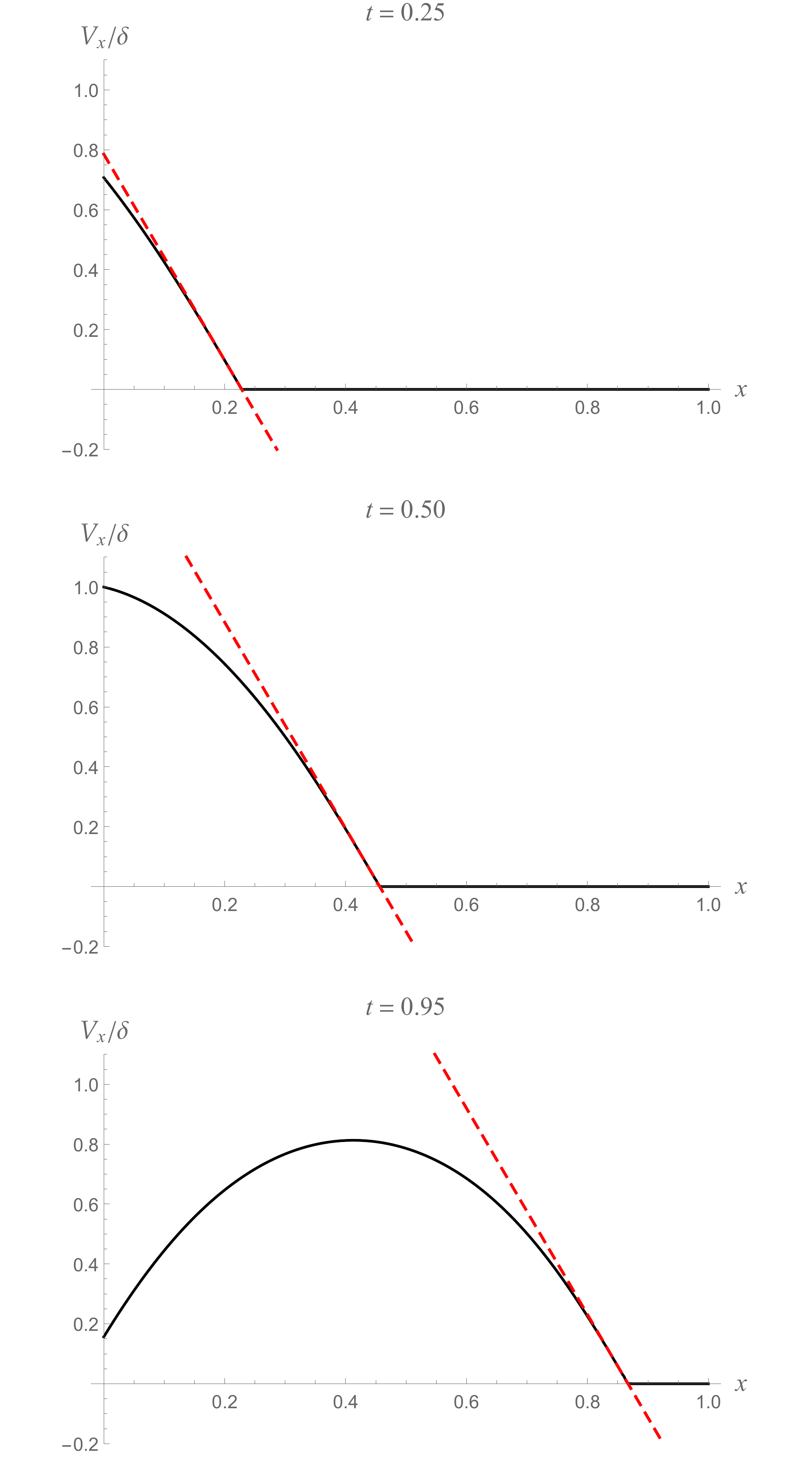, width=110mm} 
\caption{\small{$V_{x}/\delta$ vs.\ $x$ corresponding to Case~(iii)  using $\lambda =1.2$ and $\delta = -(\lambda \pi)^{-1} \approx -0.2653$, for which  $t_{\infty}= \infty$.  Black solid curves: Numerically generated  
profiles using FDS~\eqref{FDS2}.  Red broken lines: Tangents at $x=\Sigma(t)$ generated using Eq.~\eqref{Amp_IBVP}.}}
\label{fig_case_iii}  
\end{figure}
\begin{figure}[b!]
\vspace*{-20mm} 
\epsfig{file=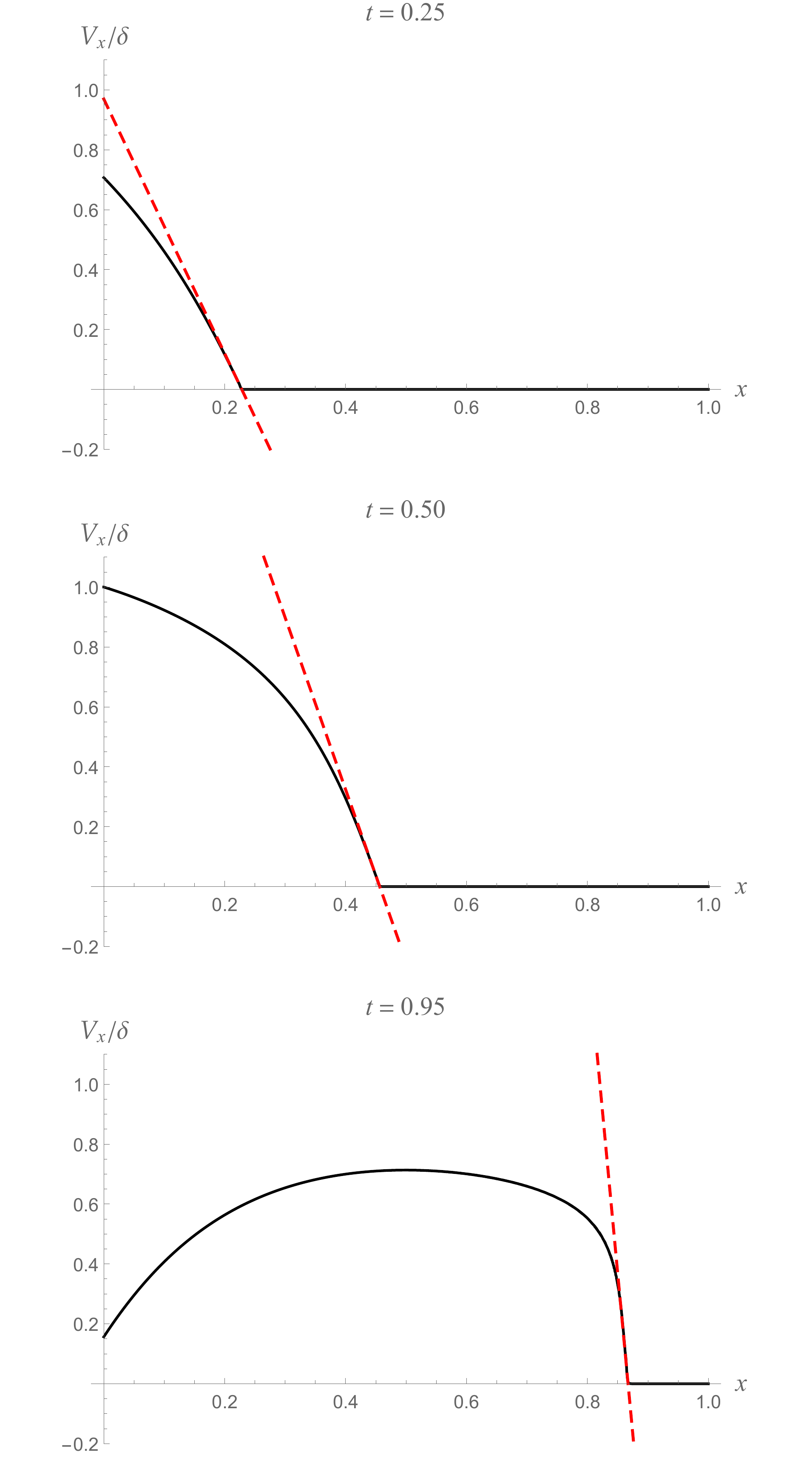, width=110mm} 
\caption{\small{$V_{x}/\delta$ vs.\ $x$ corresponding to Case~(iv)  using $\lambda =1.2$ and $\delta = -0.724$, for which  $t_{\infty}\approx 1.0951$ and $x_{\infty}\approx 0.9997$.  Black solid curves: Numerically generated  
profiles using FDS~\eqref{FDS2}.  Red broken lines: Tangents at $x=\Sigma(t)$ generated using Eq.~\eqref{Amp_IBVP}.}}
\label{fig_case_iv}  
\end{figure}


\section[Traveling wave analysis]{Traveling wave analysis\footnote{The reader 
should be aware that the analysis presented in Ref.~\cite[\S 3]{SP16} contains a 
number of omissions, misstatements, and misprints. In the present section, these 
issues have, without identification nor comment, all been remedied/corrected.}}
\label{sec:tws}

\subsection{Associated ODE, jump magnitude}
Assuming right-running waveforms propagating along the $x$-axis, we take the 
dependence of $\vartheta$ and $q$ on $x$ and $t$ to be of the form $
\vartheta(x,t)=f(\xi)$ and $q(x,t)=h(\xi)$, where $\xi :=x-vt$ is the wave variable 
and $v(>0)$ is the (constant) wave speed.  On substituting these ansatzs into  
Sys.~\eqref{Sys:GFS} we obtain, after simplifying, the system of ODEs
\begin{subequations}\label{eq_Sys_TW}
\begin{equation}\label{eq_TW_Flux}
h-v\alpha K f h'=-Kf', 
\end{equation}
\begin{equation}\label{eq_TW_Egy}
h'=v \rho c_{\rm p} f^{\prime},
\end{equation}
\end{subequations}
a (trivial) solution of which, we observe,  is  
\begin{equation}\label{eq_Sys_TW_sol1}
(h,f)= (0, \vartheta^{\bullet}).
\end{equation}
Here,   a prime denotes $\rd/\rd \xi$ and $\vartheta^{\bullet}(> 0)$ is a constant.

Now eliminating $h$ between the equations of Sys.~\eqref{eq_Sys_TW}, after  
integrating Eq.~\eqref{eq_TW_Egy} once, and  then assuming\footnote{That is, 
we are seeking kink~\cite{Angulo09}, and kink-like, traveling wave solutions.} 
that $f(\xi) \to \vartheta_{\rm r}$, $f'(\xi)\to 0$ as $\xi \to  -\infty$, we obtain the 
following Abel equation~\cite{D62} for the temperature field:
\begin{equation}\label{eq_Abel}
\kappa (1- \alpha v^2 \rho c_{\rm p} f)f'=v \vartheta_{\rm r}(1-f/\vartheta_{\rm r}),
\end{equation}
which is the associated ODE of Sys.~\eqref{Sys:GFS}.  Here, we recall that $
\kappa =K/(\rho c_{\rm p})$ is the thermal diffusivity; the constant $
\vartheta_{\rm r}$ denotes a reference  state value of $\vartheta$; and 
enforcement of the asymptotic condition gives $\mathfrak{K}_1=-\rho c_{\rm p}
v\vartheta_{\rm r}$, where $\mathfrak{K}_1$ is the resulting constant of 
integration.  

An inspection of Eq.~\eqref{eq_Abel}  reveals  that $\bar{f} = 
\vartheta_{\rm r}$, the only equilibrium point of this ODE,  is unstable for $v > 
v_{\rm a}$, but stable for $v < v_{\rm a}$, where
\begin{equation}
v_{\rm a} := \sqrt{\frac{\kappa}{\alpha \vartheta_{\rm r} K}}.
\end{equation}
 Here, we observe that  $v=v_{\rm a}$ is a degenerate case in the following 
sense:    $f= \vartheta_{\rm r}$   satisfies Eq.~\eqref{eq_Abel}, and it is also true 
that
\begin{equation}\label{eq_Abel_limit}
\lim_{f \to \vartheta_{\rm r}}\frac{v \vartheta_{\rm r}(1-f/\vartheta_{\rm r})}{\kappa 
(1- \alpha v^2 \rho c_{\rm p} f)} = \sqrt{\frac{\vartheta_{\rm r}}{\alpha \kappa K}}, 
\quad \text{but}\quad \frac{\rd f}{\rd \xi}\Bigg{|}_{f=\vartheta_{\rm r}}=0 \qquad 
(v=v_{\rm a});
\end{equation}
i.e., taking $v=v_{\rm a}$ causes $f^{\prime}$ to exhibit a jump discontinuity at $\bar{f}
=\vartheta_{\rm r}$, the magnitude of which is 
\be\label{eq:jump_TWS_dim}
\big{|} \lshad f^{\prime} \rshad \big{|} = \frac{v_{\rm a} \vartheta_{\rm r}}{\kappa}
=\sqrt{\frac{\vartheta_{\rm r}}{\alpha \kappa K}} \qquad (v=v_{\rm a}).
\en

Introducing now  the dimensionless temperature $\mathcal{T}=f/\vartheta_{\rm r}
$ and dimensionless wave variable $\eta =\xi/\ell$, Eq.~\eqref{eq_Abel} is reduced 
to
\begin{equation}\label{eq_Abel_ND}
 (1- \sigma \mathcal{T})\frac{\rd \mathcal{T}}{\rd \eta}= c (1-\mathcal{T}).
\end{equation}
Here, $\ell (>0)$ is a characteristic length; we have set
\be
\sigma := \frac{v^2}{v_{\rm a}^{2}}=\frac{\alpha v^2 \vartheta_{\rm r} K}{\kappa},
\en
 where   $\sigma =1$ implies $v=v_{\rm a}$; and  $c$, the dimensionless 
version of the wave speed  $v$,  is given by
 \be
 c = v\ell /\kappa = c_{\rm a}\sqrt{\sigma},
 \en
 where we note  that $c_{\rm a}=v_{\rm a}\ell/\kappa$ (i.e., $c_{\rm a}$ is the 
dimensionless version of $v_{\rm a}$).
 Also,  for later reference we observe that, in terms of the present dimensionless 
quantities, Eqs.~\eqref{eq_Sys_TW_sol1} and~\eqref{eq:jump_TWS_dim} 
become
\be\label{eq_Sys_TW_sol1_ND}  
 (\mathfrak{h}, \mathcal{T})=(0, \mathcal{T}^{\bullet}),
 \en
 where $\mathfrak{h}$ denotes the dimensionless version of $h$ and $
\mathcal{T}^{\bullet}=\vartheta^{\bullet}/\vartheta_{\rm r}$, and
 \be
\big{|} \lshad \rd \mathcal{T}/\rd \eta \rshad \big{|} = c_{\rm a} \qquad (\sigma=1),
\en
 respectively.
 
 \subsection{Complete stability results}
 
  A full phase plane analysis of Eq.~\eqref{eq_Abel_ND} reveals the following:
\begin{enumerate}
\item[(I)] If $\sigma >1$ and $\mathcal{T}_{\rm w}>1$, then $
\overline{\mathcal{T}}=1$ is unstable (above) and $1< \mathcal{T}(\eta) <\infty$,  
where $\mathcal{T}(\eta) \to \infty$ as $\eta \to \infty$.

\item[(II)] If $\sigma >1$ and $\sigma^{-1} < \mathcal{T}_{\rm w} <1$, then $
\overline{\mathcal{T}}=1$ is unstable (below) and $\sigma^{-1} \leq \mathcal{T}
(\eta) <1$.

\item[(III)] If $\sigma >1$ and $0 < \mathcal{T}_{\rm w} < \sigma^{-1}$, then   $-
\infty < \mathcal{T}(\eta) \leq \sigma^{-1}$, where $\mathcal{T}(\eta) \to -\infty$ 
as $\eta \to -\infty$.

\item[(IV)] If $\sigma = 1$ and $ \mathcal{T}_{\rm w} >1$, then   $1 \leq 
\mathcal{T}(\eta) < \infty$, with  $\mathcal{T}(\eta) \to \infty$ as $\eta \to \infty$.

\item[(V)] If $\sigma < 1$ and $\mathcal{T}_{\rm w}> \sigma^{-1}$, then  $
\sigma^{-1} \leq \mathcal{T}(\eta) < +\infty$, where $\mathcal{T}(\eta) \to +\infty$ 
as $\eta \to +\infty$.

\item[(VI)] If $\sigma <1$ and $1 < \mathcal{T}_{\rm w} < \sigma^{-1}$, then  $
\overline{\mathcal{T}}=1$ is stable (above) and $1 < \mathcal{T}(\eta) \leq 
\sigma^{-1}$.

\item[(VII)] If $\sigma <1$ and $0 < \mathcal{T}_{\rm w} < 1$, then $
\overline{\mathcal{T}}=1$ is stable (below) and $-\infty < \mathcal{T}(\eta) <1$, 
with $\mathcal{T}(\eta) \to -\infty, 1$ as $\eta \to \mp \infty$, respectively.
\end{enumerate}
Here, $\mathcal{T}(0)=\mathcal{T}_{\rm w}$, where  $\mathcal{T}_{\rm w}$ is a 
(known) positive constant; however, as Cases~(I), (IV), and~(VI) shall be of 
particular interest to us, we hereafter limit our attention to  $\mathcal{T}_{\rm w} 
> 1$.

Returning to Eq.~\eqref{eq_Abel_ND}, we separate variables and integrate; this 
yields, after then applying and enforcing the condition at $\eta = 0$, 
\begin{equation}\label{int_sol_bound}
c_{\rm a}\eta\sqrt{\sigma} +\mathfrak{K}_2(\mathcal{T}_{\rm w}) =
\sigma(\mathcal{T}-1)+(\sigma-1)\ln(\mathcal{T}-1) \qquad (\mathcal{T} > 1),
\end{equation} 
where $\mathfrak{K}_2(\mathcal{T}_{\rm w}) = \sigma(\mathcal{T}_{\rm w}-1)+
(\sigma-1)\ln(\mathcal{T}_{\rm w}-1)$.

In the next two subsections, the  cases of $\sigma \geq 1$ and $\sigma \in (0, 
1)$ 
are treated consecutively.


\subsection{The case $\sigma \geq 1$}
 For $\sigma \geq 1$, the integral curves take the form
\begin{equation}\label{W_sol}
\mathcal{T}(\eta) =1+
\begin{cases}
\begin{cases} 0, & \eta \in (-\infty, \eta_{\rm c}],\\
c_{\rm a} \eta+(\mathcal{T}_{\rm w}-1), &  \eta \in (\eta_{\rm c}, \infty),
\end{cases}
& \sigma =1,\\
\\
\left(\frac{\sigma - 1}{\sigma}\right)W_{0}\!\left[\left(\frac{\sigma (\mathcal{T}_{\rm 
w}-1)}{\sigma-1}\right) \exp\left(\frac{c_{\rm a} \eta\sqrt{\sigma} +\sigma (\mathcal{T}
_{\rm w}-1)}{\sigma-1} \right) \right]\!, & \sigma > 1,
\end{cases}
\end{equation}
where $W_0(\cdot)$ denotes the principal branch of the Lambert $W$-
function~\cite{Corless96} and we have set $\eta_{\rm c} := c_{\rm a}^{-1}(1-
\mathcal{T}_{\rm w})$. For the case  $\sigma = 1$, the solution profile is  seen to 
be a piecewise-linear, but continuous,  function of $\eta$; it was constructed by 
joining together, at the point  $\eta = \eta_{\rm c}$,  
Eq.~\eqref{eq_Sys_TW_sol1_ND}, with    $\mathcal{T}^{\bullet} = 1$, and the $
\sigma = 1$ special case of Eq.~\eqref{int_sol_bound}.

\begin{center}
\begin{figure}[t!]
\epsfig{file=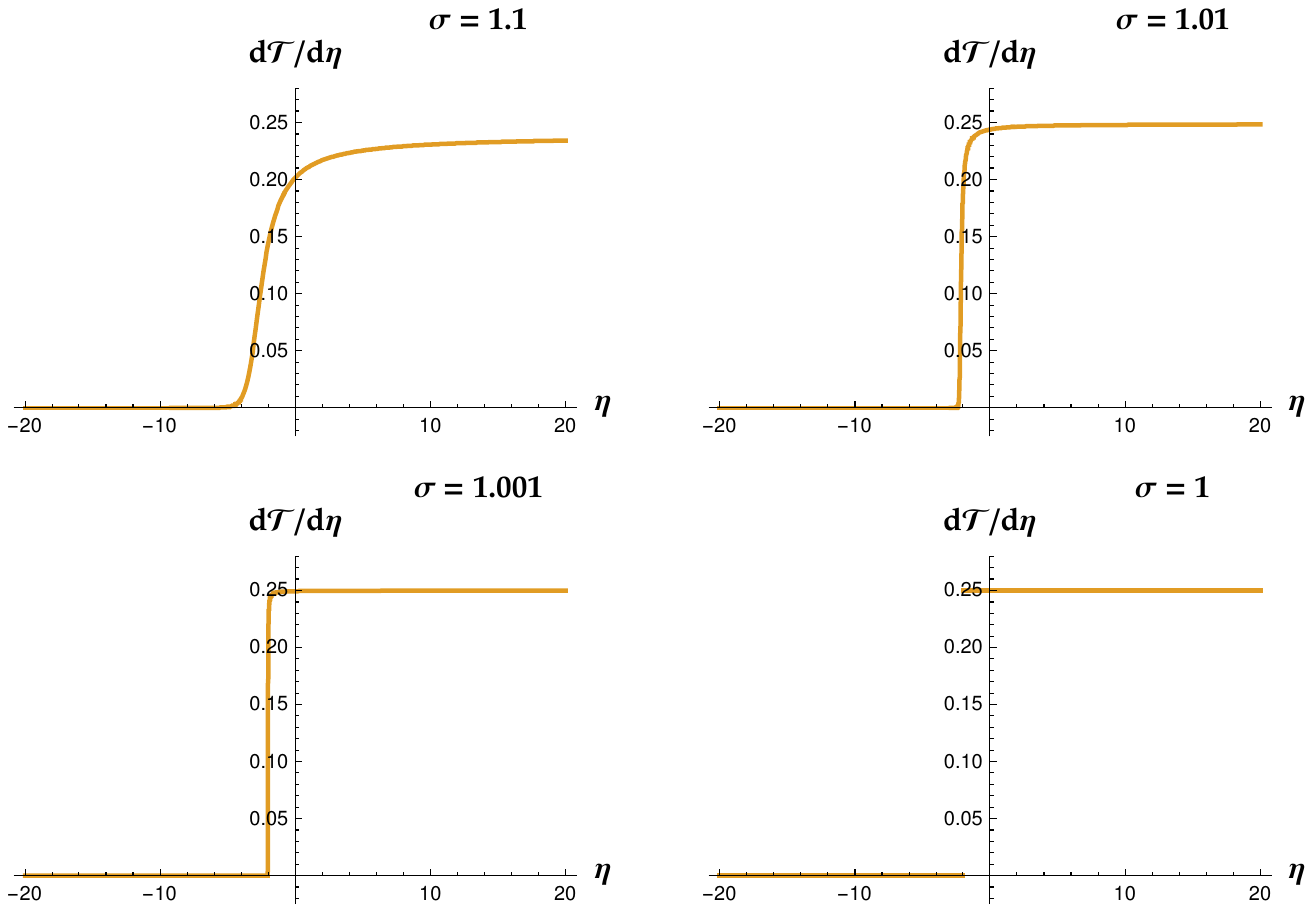, width=160mm} 
\vspace*{-122mm}  
\caption{\small{$\rd \mathcal{T}/\rd \eta$ vs.\ $\eta$, generated from 
Eq.~\eqref{eq:Accel-TWS}, for  $c_{\rm a}=0.25$  and $\mathcal{T}_{\rm w}=1.5$.  
Here, $\eta_{\rm c}=-2$.}}
\label{fig_Accel-TWS}  
\end{figure}
\end{center}

From Eq.~\eqref{W_sol} we find that  $\mathcal{T}\! \to \!\infty$ as $\eta \!
\to \!\infty$.  It can, however, be shown that the temperature gradient in this case 
is bounded; specifically, \hbox{$0 \leq \rd \mathcal{T}(\eta)/\rd \eta \leq c_{\rm a}$}, for all $
\sigma \geq 1$, where  in this subsection the temperature gradient is given by
\begin{equation}\label{eq:Accel-TWS}
\frac{\rd \mathcal{T}(\eta)}{\rd \eta}=
\begin{cases}
\begin{cases} 0, &  \eta \in (-\infty, \eta_{\rm c}],\\
c_{\rm a}, & \eta \in (\eta_{\rm c}, \infty),
\end{cases}
& \sigma =1,\\
\\
\displaystyle{\frac{c_{\rm a}}{\sqrt{\sigma}}\left\{\frac{ W_{0}\!\left[\left(\frac{\sigma 
(\mathcal{T}_{\rm w}-1)}{\sigma-1}\right) \exp\left(\frac{c_{\rm a}\eta\sqrt{\sigma} +
\sigma (\mathcal{T}_{\rm w}-1)}{\sigma-1} \right) \right]}{ 1+W_{0}\!
\left[\left(\frac{\sigma (\mathcal{T}_{\rm w}-1)}{\sigma-1}\right) \exp\left(\frac{c_{\rm a} 
\eta\sqrt{\sigma} + \sigma (\mathcal{T}_{\rm w}-1)}{\sigma-1} \right) \right]}
\right\}}, & \sigma > 1.
\end{cases}
\end{equation}
From Eq.~\eqref{eq:Accel-TWS} it is clear that $\sigma =1$ corresponds to 
Case~(iii), i.e., the constant amplitude case, of our temperature-rate wave 
analysis (recall Sect.~\ref{sect:accel}).  The sequence shown in  Fig.~\ref{fig_Accel-TWS}  
depicts the steepening of the temperature gradient profile as $\sigma \to 1$ 
(from above); in this limit, the $\rd \mathcal{T}(\eta)/\rd \eta$ vs.\ $\eta$ profile 
tends to a step function, i.e., a temperature-rate wave in the present setting, of 
(jump) magnitude $c_{\rm a}$.

\begin{center}
\begin{figure}[t!]
\epsfig{file=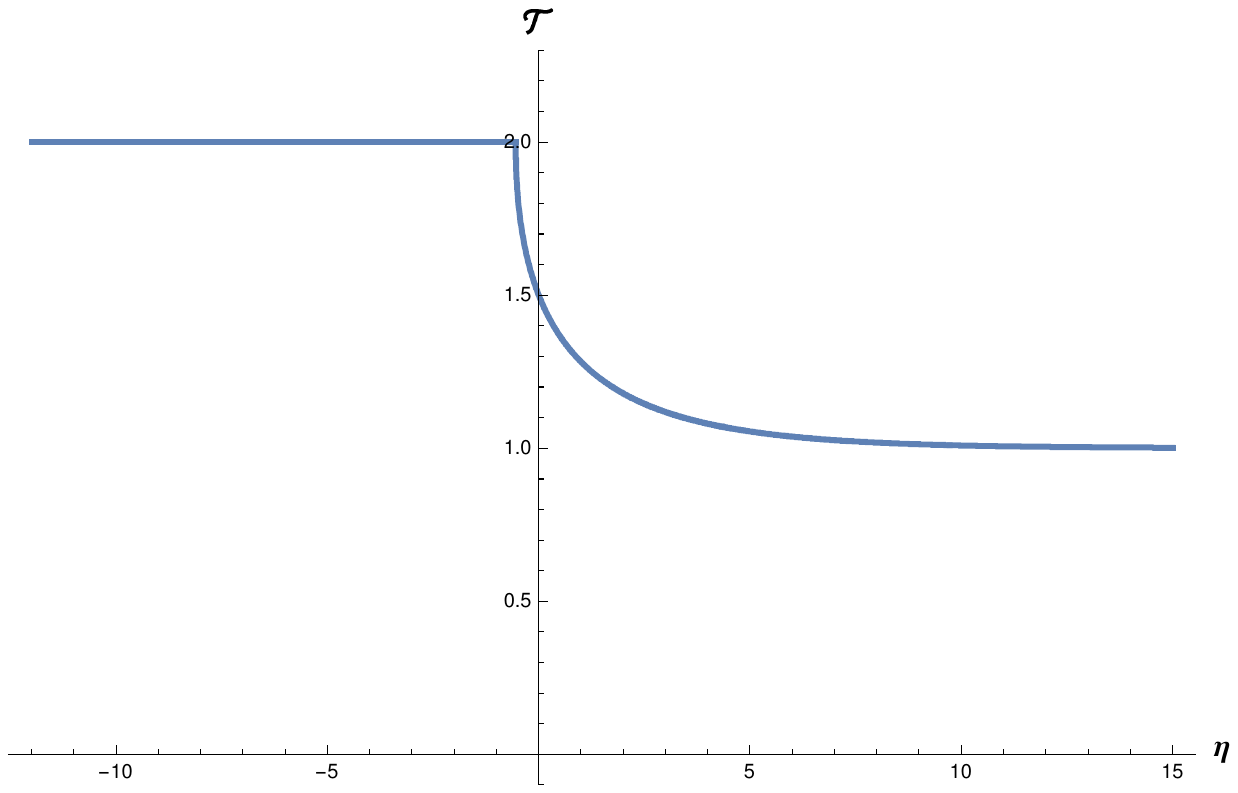, width=150mm} 
\vspace*{-120mm}  
\caption{\small{$\mathcal{T}$ vs.\ $\eta$, generated from 
Eq.~\eqref{W_sol_bound}, for  $\sigma = 0.5$, $c_{\rm a}=0.25$,  and $\mathcal{T}
_{\rm w}=1.5$. Here, $c \approx 0.1768$ and $\eta^{*}\approx -0.5463$.}}
\label{fig_Temp-TWS}  
\end{figure}
\end{center}

\subsection{The case $\sigma \in (0, 1)$}
On joining, at the point $\eta=\eta^{*}$ (see below), the $\sigma \in (0, 1)$ 
special case of Eq.~\eqref{int_sol_bound} to  the constant-temperature solution 
given in Eq.~\eqref{eq_Sys_TW_sol1_ND},  the piecewise-defined integral 
curve corresponding to this range of $\sigma$-values is readily constructed, viz.: 
\begin{multline}\label{W_sol_bound}
\mathcal{T}(\eta) =1+\sigma^{-1}\\
\times \!\begin{cases}
1-\sigma, & \eta \leq \eta^{*}\\
\\
(\sigma - 1)W_{0}\!\left[\left(\frac{\sigma (\mathcal{T}_{\rm w}-1)}{\sigma-1}\right) 
\exp\left(\frac{c_{\rm a}\eta\sqrt{\sigma} +\sigma (\mathcal{T}_{\rm w}-1)}{\sigma-1} 
\right) \right]\!, &  \eta > \eta^{*}
\end{cases}
\quad (\sigma < 1).
\end{multline}
Here, we note the restriction $\mathcal{T}_{\rm w} \in (1, \sigma^{-1})$ and  
observe that    the value of $\mathcal{T}(\eta)$ on the interval $\eta\leq \eta^{*}$ 
follows on setting  $\mathcal{T}^{\bullet} = 1/\sigma$ in 
Eq.~\eqref{eq_Sys_TW_sol1_ND}.   Also, $\eta^{*}(< 0)$ is given by
\begin{equation}\label{xi_astk}
\eta^{*} := \frac{1}{c_{\rm a}\sqrt{\sigma}}\left\{ 1-\sigma \mathcal{T}_{\rm w}-(1-
\sigma) \ln \left [\frac{1-\sigma}{\sigma(\mathcal{T}_{\rm w}-1)} \right] \right\}\!,
\end{equation}
where Eq.~\eqref{xi_astk} was obtained by setting the argument of $W_0$ (in 
Eq.~\eqref{W_sol_bound}) equal to $-1/\re$ and then solving for  $\eta$, where 
$-1/\re$ is a branch point of the $W$-function; again, see 
Ref.~\cite{Corless96}.

Along with the fact that Eq.~\eqref{W_sol_bound} describes a bounded, 
continuous, waveform, for which $\overline{\mathcal{T}}=1$ is a stable 
equilibrium,   Fig.~\ref{fig_Temp-TWS} also reveals that for $0< \sigma <1$, the 
GFS traveling wave profile exhibits what Zel'dovich and Raizer~\cite[Fig.~10.3b]{ZR02} refer to as  a (preheating)  `tongue'.

\section{Closure}\label{sec:concl}

From the mathematical standpoint, the present analysis has shown that the qualitative 
behavior of  Sys.~\eqref{Sys:GFS} is very much like that exhibited by the   1D version of the 
`Darcy--Jordan' 
model\footnote{Also known as the `Jordan--Darcy' model; see, e.g., 
Ref.~\cite{S08} and those cited therein.}~\cite{j05} (of poroacoustics)  when the fluid phase consists of a  
\emph{retrograde fluid}\footnote{Fluids that exhibit rarefaction (or `negative') shocks; see, e.g., 
Ref.~\cite{T91} and those cited therein.  Note also that retrograde fluids correspond to  $\beta <0$ in 
Ref.~\cite{j05}, wherein $\beta$  denotes the coefficient of nonlinearity.}.  This analogy is most evident
in the case of temperature-rate waves;  recall the behavior of $U_{\rm wnl}(\vartheta)$ (see 
Sect.~\ref{sect:MathPrelim}),  and observe
that $-V_{x}$ corresponds to $p^{\prime}$, where $p^{\prime}$ is used in Ref.~\cite[\S 4]{j05} to denote the 
dimensionless over pressure in a regular fluid.

It is also noteworthy that the  behavior of the waveforms observed in 
Sect.~\ref{sec:tws} under the $\sigma \geq 1$ and $\sigma \in (0,1)$ cases 
corresponds to taking $\epsilon >  0$ and $\epsilon < 0$, respectively, in 
Christov and Jordan~\cite[pp.~1126--1128]{cj10}, who examined traveling waves 
under the MC law with  $\mathcal{K}(\vartheta)$ a linear function of $\vartheta$ 
and $\tau(\vartheta) :=$~const. 

With regard to possible follow-on studies, the obvious next step from the 
numerical standpoint is to examine thermal shock phenomena under 
Sys.~\eqref{Sys:GFS} using what are known as `shock capturing' schemes (see, 
e.g., Ref.~\cite{cj10} and those cited therein), which are more elaborate than the  
simple explicit scheme we employed in Sect.~\ref{sec:numerical}. From the 
analytical standpoint, extensions of the present  study might include performing the above 
analyzes on the GFS special cases of MR theory and CFO theory; recall the second and 
third bulleted items in Sect.~\ref{sect:Intro}.

On the other hand, Sys.~\eqref{Sys:GFS_3D} could be recast in terms of what 
Straughan \cite{S10} has termed  `Cattaneo--Christov' theory.  Under this 
generalization of the MC law, which Christov \cite{CIC09} proposed in 2009, the 
simple partial time derivative that acts on ${\bf q}$ would be replaced by a Lie 
derivative---one corresponding to Oldroyd's upper convected derivative---which would then make Sys.~\eqref{Sys:GFS_3D} applicable to \emph{moving} (GFS) conductors; to this latter point, see also the first footnote in Ref.~\cite[p.~136]{CFO82}.

\section*{Acknowledgments}
The authors are grateful  to the anonymous referee for his/her helpful comments and  for bringing 
Ref.~\cite{ColLai94} to their attention.   S.C.\ thanks  the financial support of G.N.F.M.--I.N.d.A.M.,  
I.N.F.N.\ and Universit\`a di Roma  \textsc{La Sapienza}, Rome, Italy.

\end{document}